\documentclass[%
reprint,
amsmath,amssymb,
aps,
pra,
notitlepage
]{revtex4-2}% preamble:
\usepackage[english]{babel}
\usepackage[utf8]{inputenc}
\usepackage{amsmath}
\usepackage{physics}
\usepackage{graphicx}
\graphicspath{{figures/}}
\usepackage[colorinlistoftodos]{todonotes}
\usepackage{lipsum}
\usepackage{multirow}
\usepackage{calc}
\usepackage[bottom]{footmisc}
\usepackage[]{hyperref}
\usepackage{booktabs}
\usepackage[normalem]{ulem}
\usepackage{float}
\usepackage{lineno}

\newcommand{\figref}[1]{Fig.~\ref{#1}}

\begin{document}
\title{Low random duty-cycle errors in periodically-poled KTP revealed by sum-frequency generation}

\author{Felix Mann$^{1,\star}$, Helen M. Chrzanowski$^{1}$, and Sven Ramelow$^{1}$}
\affiliation{$^{1}$Institut f\"ur Physik, Humboldt-Universit\"at zu Berlin, Newtonstr. 15, 12489 Berlin, Germany \\
$^{\star}$Corresponding author: felixmann@physik.hu-berlin.de}

%% To be edited by editor
% \dates{Compiled \today}

%\ociscodes{(140.3490) Lasers, distributed feedback; (060.2420) Fibers, polarization-maintaining;(060.3735) Fiber Bragg gratings.}

%% To be edited by editor
% \doi{\url{http://dx.doi.org/10.1364/XX.XX.XXXXXX}}

\begin{abstract}
Low-noise quantum frequency conversion in periodically-poled nonlinear crystals has proved challenging when the pump wavelength is shorter than the target wavelength. This is -- at least in large part -- a consequence of the parasitic spontaneous parametric down-conversion of pump photons, whose efficiency is increased by fabrication errors in the periodic poling. Here we characterise the poling quality of commercial periodically-poled bulk potassium titanyl phosphate (ppKTP) by measuring the sum-frequency generation (SFG) efficiency over a large phase mismatch range from 0 to more than 400$\pi$. Over the probed range the SFG efficiency behaves nearly ideally and drops to a normalised efficiency of $10^{-6}$. Our results demonstrate that any background pedestal which would be formed by random duty cycle errors in ppKTP is substantially reduced when compared to periodically poled lithium niobate. The standard deviation of the random duty cycle errors can be estimated to be smaller than 2\% of the domain length. From this, we expect a noise spectral density which is at least one order of magnitude smaller than that of current state-of-the-art single-step frequency converters. 
 
\end{abstract}

%\setboolean{displaycopyright}{true}

\maketitle
The realisation of quantum networks will require hybrid quantum technology, demanding efficient inter-conversion of stationary and flying qubits \cite{Kumar:90,Ou2008,Kimble:2008if,Wehner18}. An indispensable ingredient to its long range implementation will be quantum frequency conversion (QFC) \cite{Kumar:90}, allowing photons at single quantum emitter wavelengths to interface with low loss telecommunication infrastructure \cite{MATALONI1998,Albota04,Langrock05,Tanzilli05,VanDevender07, rakher10, Takesue10, zaske12,dreau}. For example, 637.5 nm photons from nitrogen-vacancy (NV) centers in diamond -- a leading candidate for the implementation of quantum networks -- \cite{diamond,Kimble:2008if} would suffer dramatic transmission losses without their conversion to the telecom band \cite{dreau}. Although four-wave-mixing can also be used \cite{rakher10,McKinstrie12,Li:2016eg} for implementing QFC, it is predominantly realised in periodically-poled $\chi^{(2)}$-nonlinear crystals using either sum-frequency or difference-frequency generation (SFG/DFG).

So far -- and with a handful of exceptions -- waveguided periodically poled lithium niobate (ppLN) has been the favoured workhorse for QFC implementations \cite{Langrock05,Tanzilli05,Takesue10,Curtz2010, zaske12,Ates2012,Fernandez2013,ikuta,dreau,Bock2018,Walker2018,fejer2step,maring,strassmann}; especially so for the specific task of converting 637.5 nm photons to the telecom band \cite{dreau,pelcthesis,fejer2step,ikuta,maring,strassmann}. Its broad availability and large nonlinear coefficient, alongside the growing maturity of waveguide fabrication and relatively low pump power requirements underlie its popularity for this task. However, ppLN-based realisations of single-stage conversion with a pump wavelength shorter than at least one of the involved single-photon wavelengths, suffer from background noise that strongly limits the performance of conversion (see Table \ref{tab:comp}). This noise background can emerge from spurious processes or fabrication imperfections, including waveguide non-uniformity, {\v{C}}erenkov-idler radiation \cite{rastogi}, Raman processes \cite{Kuo2013}  and notably also for this work, random duty cycle (RDC) errors \cite{fejer92,pedestal,phillips}. This has resulted in the conclusion that a two-step conversion process -- moving the required pump to a much longer wavelength and thus omitting spurious SPDC fully -- would be the preferable option for low noise QFC from 637.5 nm to the telecom band \cite{fejer2step}.

\begin{figure}[!htbp]
\centering
\includegraphics[width=9cm]{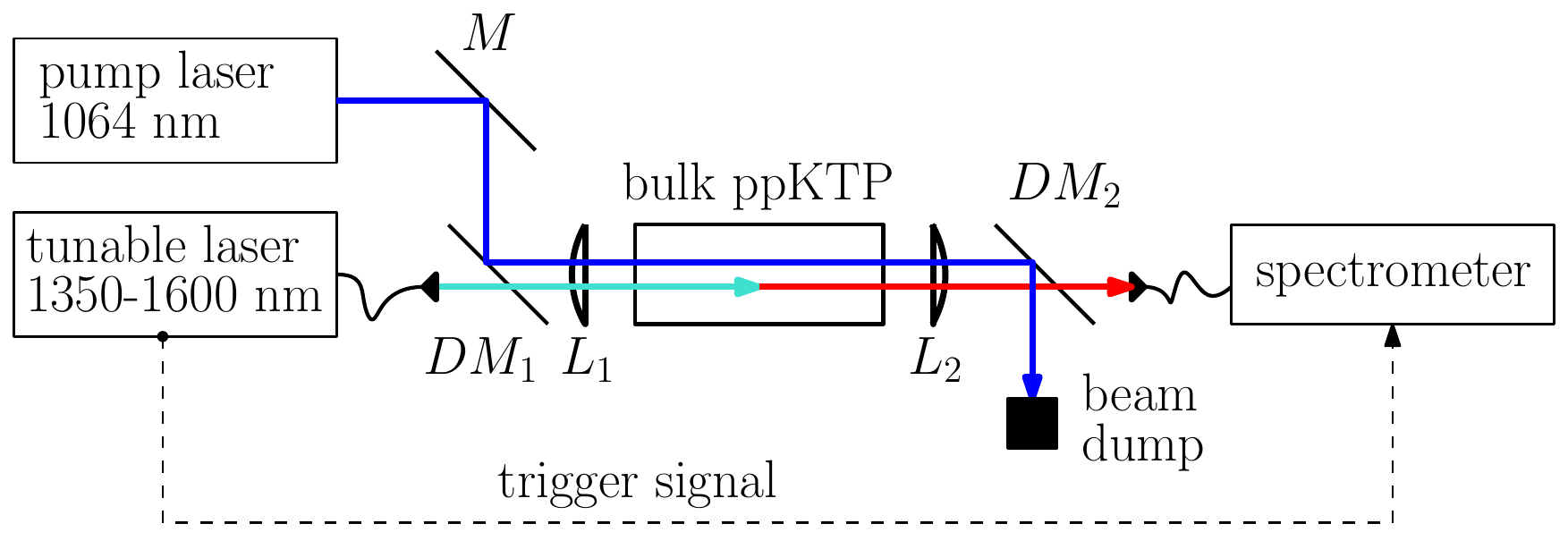}
\caption{RDC error characterisation based on SFG: A widely tunable laser and 1064 nm pump are focused into a ppKTP crystal, while the SFG output is measured by a sensitive spectrometer.}
\label{fig:setup}
\end{figure}

\begin{table*}[!htbp]
\centering
\caption{\bf Comparison between the generated noise spectral densities (NSD) for similar conversion experiments in ppLN waveguides and our approach based on bulk ppKTP}
\begin{tabular}{p{5cm}p{1.7cm}p{2cm}p{2cm}p{2.5cm}p{2.5cm}}
\hline \hline
& Detection bandwidth& Measured noise& Measured  NSD&$\eta_{filter} \, \eta_{fc} \, \eta_{detector}$ &  Generated NSD \\
\hline \hline
Esfandyarpour et al. \cite{fejer2step} , \textit{ppLN WG} & - & - & - & - & 0.24 kHz/nm \\
\hline
Dr{\'e}au et al.\cite{dreau}, \textit{ppLN WG}  & 0.004 nm & 0.27 kcts/s & 0.07 MHz/nm & $0.41\cdot 0.85 \cdot 0.41$ & 0.5 MHz/nm  \\
Pelc et al.\cite{NSD} , \textit{ppLN WG} & 44 nm & - & - & - & 1 MHz/nm  \\
Ikuta et al. \cite{ikuta}, \textit{ppLN WG}   & 0.7 nm & 45 kcts/s & 0.07 MHz/nm & $0.31\cdot 0.083 \cdot 0.6$  & 4.2 MHz/nm \\
Strassmann et al. \cite{strassmann}, \textit{ppLN WG}   & 0.2 nm & 100 kcts/s & 0.5 MHz/nm &  $0.4\cdot 0.75 \cdot 0.1$ & 17 MHz/nm \\
Maring et al. \cite{maring} , \textit{ppLN WG} & 0.008 nm & 35 kcts/s & 4.4 MHz/nm & $0.62\cdot 0.79 \cdot 0.1$ & 90 MHz/nm \\
\hline
here, \textit{bulk ppKTP} & - & - & - & - & $< 0.02$ MHz/nm  \\
\hline
\end{tabular}
  \label{tab:comp}
\end{table*}

While some of the noise processes that constrain QFC in ppLN waveguides are absent when using bulk ppLN, RDC errors remain a significant bottleneck. However, to minimise random duty-cycle errors and therefore mitigate this noise source, another commonly used nonlinear crystal appears to be a promising candidate: periodically poled potassium titanyl phosphate (ppKTP). In the past decades ppKTP has become the material of choice for numerous nonlinear optical applications, owing to its relatively large nonlinearity (on par with that of LN), broad transparency, high-damage threshold and most importantly its suitability for quasi-phase matching. 

Significant here, however, is KTP's coercive field, which is about an order of magnitude less than that required for LN, and its highly anisotropic ferroelectric domain propagation velocities along the different crystal axes, which significantly limits domain broadening \cite{Urenski01,Canalias05}. This permits the manufacture of dense sub-micron poling periods in ppKTP crystals \cite{Canalias07,Zukauskas11} and is evidence for much more regular domain structures than those currently available in LN. 

Here, we experimentally test the hypothesis that commercially available ppKTP crystals have comparatively low random duty-cycle errors. We implement an extension of the technique introduced in \cite{pedestal} that relies on precise measurements of the SFG efficiency over large detunings of $\Delta k$ and a corresponding dynamic range of more than 6 orders of magnitude. We show that for a commercial ppKTP crystal $\overline{\sigma}$ (the standard deviation of of the RDC errors) is at most 2\%, which would result in noise spectral density (NSD) values for maximum efficiency QFC at least one order of magnitude below the previously reported state-of-the-art in ppLN. These results demonstrate that bulk ppKTP is potentially a superior architecture for frequency conversion and perhaps other applications that would benefit from low RDC errors.

In quantum frequency conversion, for sufficiently large pump powers, the conversion efficiency $\eta_c$ can theoretically reach 100\%. The conversion efficiency as a function of pump power, $P_P$ is given as \cite{Albota04}
\begin{gather}
 \eta_c = \sin^2\left( \frac{\pi}{2}\sqrt{\frac{P_P}{P_{max}}}\right),\hspace{0.3cm}P_{max} = \frac{c \epsilon_0 n_t n_r \lambda_t \lambda_r \lambda_p}{128d^2_{eff}L h_m}.
 \label{eq:SFG}
\end{gather}
where $\lambda _x$ is the vacuum wavelength with the subscripts denoting $t$ for target or telecom, $r$ for red and $p$ for pump (with $n_x$ denoting their corresponding refractive indices). $d_{eff}$ is the nonlinear coefficient, $L$ is the length of the converter and $h_m$ the reduction factor for focused Gaussian beams, which is a function of the confocal parameters of the participating beams. Typically, efficient conversion demands large pump powers corresponding to up to $10^{20}$ pump photons per second. Thus, on the single photon level, even very weak scattering processes of the pump can produce a strong background noise at the target wavelength that can strongly contaminate the conversion process \cite{dreau,Albota04,ikuta,NSD}. This is especially relevant for scenarios where the pump wavelength is shorter than one of the single-photon wavelengths. For converting photons from NV centers with a wavelength of 637.5 nm ($\lambda_r$) to the telecom wavelength $\lambda_t$ via the assistance of a strong pump at $\lambda_p= 1064.5\mbox{ nm}$, the dominant pump-induced noise contribution stems from spurious spontaneous parametric downconversion (SPDC). This parasitic background is vastly enhanced by random duty-cycle (RDC) errors caused by imperfections in the periodic poling process \cite{fejer92,pedestal}. These fabrication errors introduce a pedestal in the otherwise quadratically decreasing SPDC efficiency far from phase matching \cite{pedestal}. The efficiency of the noise process for an ensemble average of many gratings $\langle\eta_{noise}\rangle$ as a function of the phase mismatch $\Delta k=\Delta k_{noise}-\Delta k_{poling}$ is given by \cite{pedestal}
\begin{gather}
   \langle  \eta_{noise}(\Delta k) \rangle =e^{-\frac{\pi^2\overline{\sigma}^2}{2}}\cdot\mbox{sinc}^2\left(\frac{\Delta k L}{2}\right)+\frac{1-e^{-\frac{\pi^2\overline{\sigma}^2}{2}}}{N_D}.
     \label{eq:pedestal}
\end{gather} Here $\overline{\sigma}=\frac{\sigma}{l}$ is the standard deviation of of the RDC errors per domain length $l$ and $N_D=\frac{L}{l}$ the total number of poled domains. The domain length is given by $l=\frac{\Lambda}{2}$ where $\Lambda$ is the poling period. Far from the phase matching point (for large $\Delta k$) the second term in \eqref{eq:pedestal} dominates and forms the aforementioned pedestal. It has been shown in \cite{pedestal} that one can infer $\overline{\sigma}$ from the pedestal height using \eqref{eq:pedestal}, as confirmed by independent measurements of $\overline{\sigma}$ with a Zygo interferometer\cite{pedestal}. Using the expression for the SPDC rate from \cite{spdc}, \eqref{eq:SFG} and \eqref{eq:pedestal} the noise spectral density (NSD) for small $\overline{\sigma}$ and plane wave interactions can be estimated as \cite{NSD}
\begin{gather}
    NSD=\langle \eta_{noise}\rangle\cdot\frac{dN_t}{d\lambda_t}\approx\frac{\pi^2\overline{\sigma}^2}{N_D}\cdot \frac{n_r}{n_m}\frac{\lambda_rc}{2\lambda_m\lambda_t^2}\cdot \arcsin(\sqrt{\eta_c}),
    \label{eq:NSD}
\end{gather}
where $\lambda_m$ is the idler wavelength of the noise process. Note that the NSD is independent of the necessary pump power to achieve $\eta_c=1$.
\begin{figure*}[!htbp]
\centering
\includegraphics[width=0.85\textwidth]{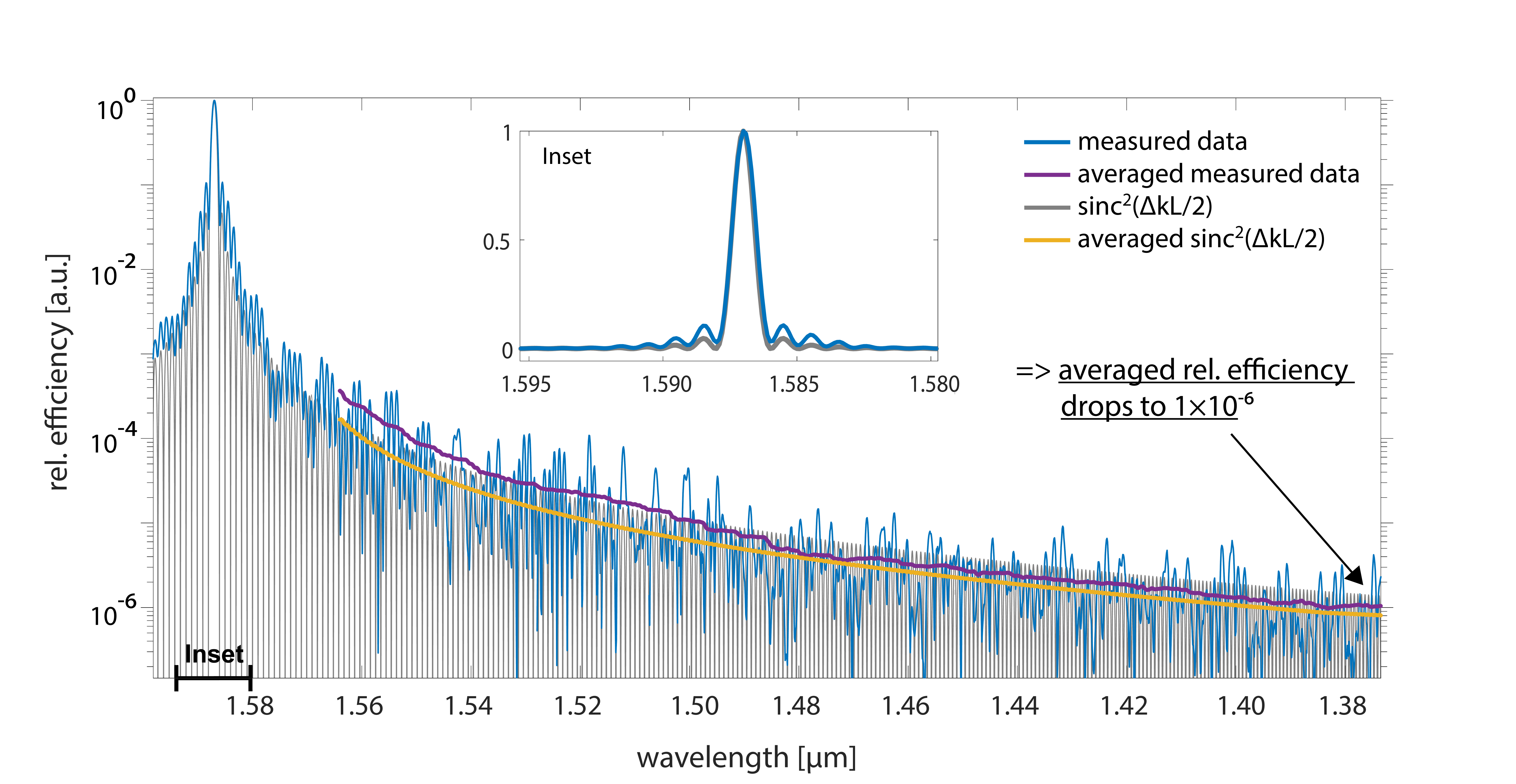}
\caption{Experimentally measured SFG efficiency (blue) as a function of input wavelength together with the theoretical expectation (grey) for an ideal poling $sinc^2(\Delta k L/2)$ . The purple curve is the moving average (30nm averaging window) of the measured data and the yellow curve the moving average of the $sinc^2$. Inset: linear scale SFG-efficiency near the phase matching peak.}
\label{fig:m}
\end{figure*}
In order to bound the magnitude of the RDC errors present in our standard, commercially-sourced KTP crystal, we realised a SFG process that allows us to measure the SFG efficiency at very large detunings from $\Delta k = 0$. The experimental setup is presented in \figref{fig:setup}. A 20 mm long KTP crystal (Tailored Photons) is quasi-phase matched for type-0 SFG(DFG) process 1589 nm +  1064.5 nm $\rightarrow$ 637.5 nm (or vice versa) and has a domain length of $l=7.85\mbox{ }\mu$m and $N_D\approx 2550$. The conversion bandwidth is 110 GHz equating to 0.9 nm (0.15 nm) at 1589 nm (637.5 nm). For the characterisation, a 1064.5 nm Nd:YAG pump and step-tunable input lasers were used. The wavelength scan used two tunable telecom lasers to cover the L- and C-telecom bands. The pump and input beams are superimposed and focused into the crystal with waists of 35 $\mu$m and 50 $\mu$m respectively, yielding equal Rayleigh ranges of about 6.5 mm. The red light generated in the crystal via SFG is subsequently detected via a grating spectrometer with a sensitive back-thinned CCD sensor.
To scan the phase matching curve, the input laser was step-wise tuned from 1600 nm to 1370 nm in 0.1 nm steps. To drive the SFG process 260 $\mu$W and 40 mW of input and pump power were used respectively. The resulting SFG signal, detected via the spectrometer, was determined by integrating the spectrum over a 20 nm window around the peak. An unwanted background signal from SFG of the amplified spontaneous emission (ASE) of the input laser at 1589 nm was filtered out by two coarse wavelength division multiplexers with a center wavelength of 1591 nm. 
\par
\begin{figure}[!htbp]
\centering
\includegraphics[width=0.9\columnwidth]{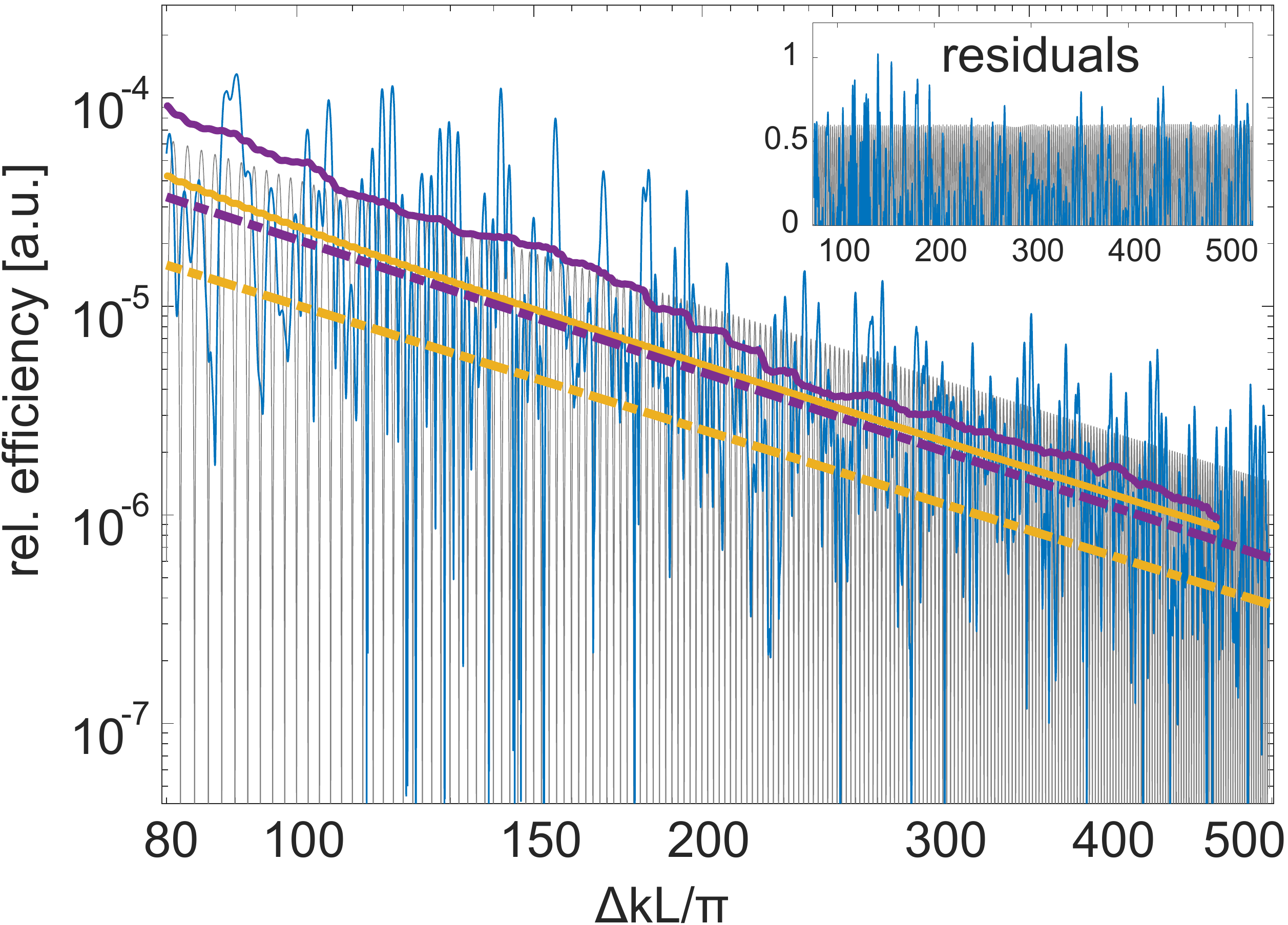}
\caption{Double-logarithmic plot of the SFG efficiency far from the peak. blue: measured SFG efficiency; grey: ideal theoretical $sinc^2$ roll-off; purple (yellow) dash-dotted: linear fit of the measured data (theory) in double-logarithmic coordinates; purple (yellow) solid lines: moving average of the measured data (theory); inset: fit residuals of both linear fits.}
\label{fig:loglog}
\end{figure}

 The measured SFG efficiency, normalised to the peak value, is shown in \figref{fig:m}. Across the measured range we observe no pedestal. The periodic modulation, evident in \figref{fig:m} and \figref{fig:loglog}, with a period of about 5 nm is attributed to an etaloning effect in the backthinned CCD of the spectrometer. The presence of this etaloning effect was independently verified with a tunable laser over 635-640 nm. For values of $\tfrac{\Delta k L}{\pi}< 200$ our average measured SFG efficiency sits slightly above the anticipated theoretical curve (e.g. Fig. \ref{fig:loglog}), indicating that more complicated poling errors could be present and contribute to an increase in SFG across smaller $\Delta k$. However for larger $\Delta k$ the curve behaves more ideally again.

\par
\begin{figure}[!htbp]
\centering
\includegraphics[width=0.9\columnwidth]{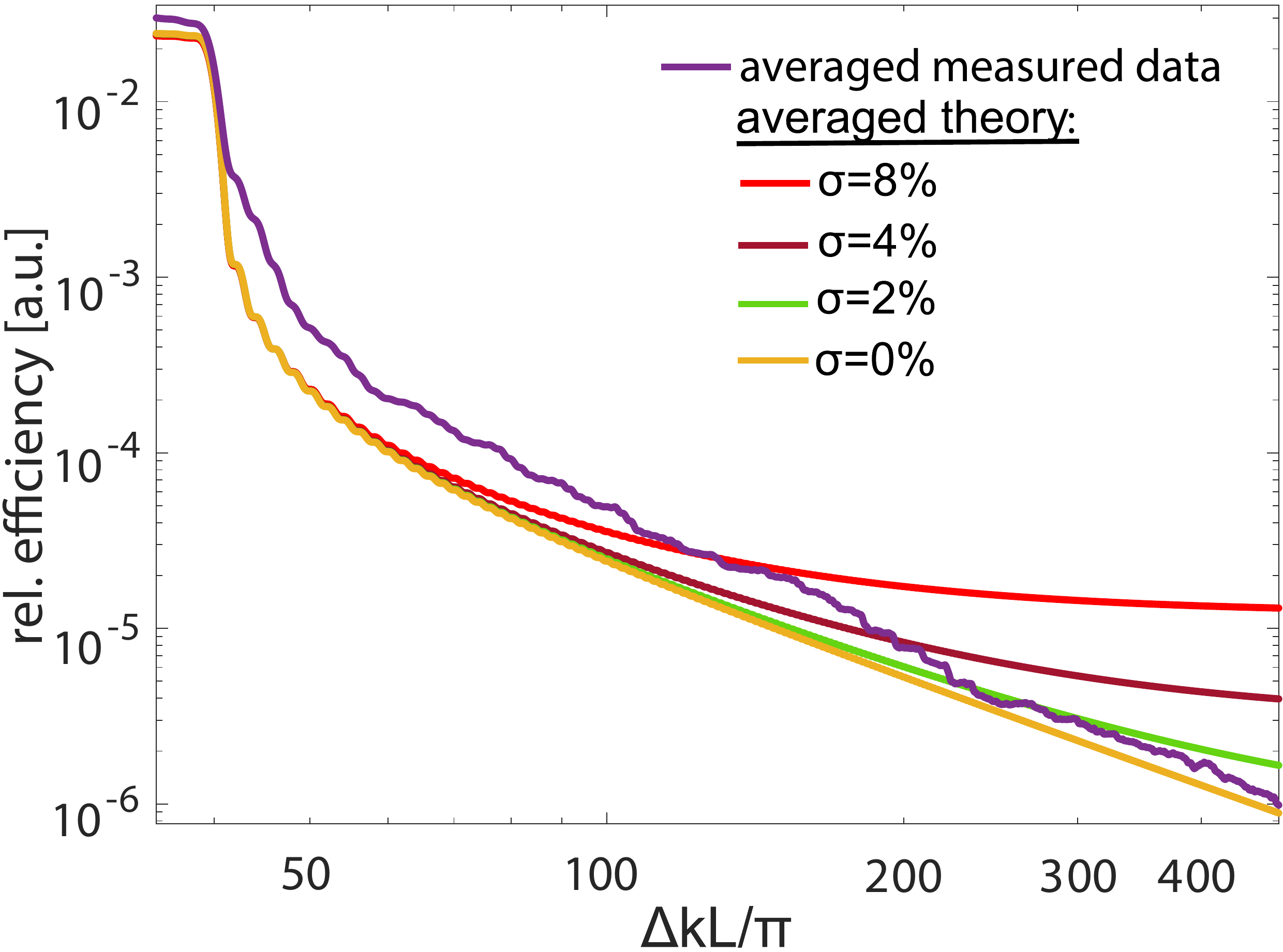}
\caption{Double-logarithmic plot for moving average of the measured SFG efficiency (purple) and the predicted efficiencies (\eqref{eq:pedestal}) for different RDC errors: $\overline{\sigma}=0\%$ (yellow), $\overline{\sigma}=2\%$ (green), $\overline{\sigma}=4\%$ (brown) and $\overline{\sigma}=8\%$ (red) using a constant averaging window of $\Delta k L/\pi= 80$.}
\label{fig:movmeans}
\end{figure}

 The SFG efficiency far from $\Delta k = 0$ as a function of $\Delta k L/\pi$ is shown in \figref{fig:loglog}. The double-logarithmic representation reveals the expected quadratic efficiency decrease. Linear fits of both the theory and measured data indeed yield a slope very close to -2. The fit residuals are shown in the inset \figref{fig:loglog}. Evidence of a pedestal would produce an increase in the residuals for increasing detuning, which is not evident over the probed range. From \eqref{eq:pedestal} we can calculate an upper bound for the RDC error of $\overline{\sigma}=2 \%$, based on absence of a pedestal in the averaged efficiency curve above $1 \cdot 10^{-6}$. In \figref{fig:movmeans} the moving average of the measured data and the predicted behaviour for different RDC errors are plotted. It confirms that our measured efficiency drops well below the $\overline{\sigma}=2\%$ prediction. In comparison, the best measured $\overline{\sigma}$ for lithium niobate waveguides is 8\% \cite{pedestal}, while the average $\overline{\sigma}$ is specified as about 10\% \cite{pelcthesis}. In an experiment with similar wavelength combination to that analysed here, an `effective' $\overline{\sigma}$ of 21\% was inferred \cite{NSD}, capturing both the effect of RDC errors in the waveguides and {\v{C}}erenkov-idler radiation. The aforementioned problem of unguided modes due to {\v{C}}erenkov-idler radiation and the resulting increase of converter bandwidth is a problem unique to waveguides and does not arise in a bulk crystal. Nevertheless, the $\overline{\sigma}$ for a bulk ppLN implementation \cite{Albota04} was inferred to be comparable to the 28\% of \cite{NSD}.
 
To ascertain the potential improvements to QFC afforded by bulk ppKTP we calculate an expected NSD for a maximum conversion efficiency. Using our upper bound of 2\% RDC errors and \eqref{eq:NSD} we obtain about 20 kHz/nm. This result is compared in Table \ref{tab:comp} to similar conversion experiments: Apart from the two-step conversion scheme realised by \cite{fejer2step}, the NSD for the bulk ppKTP platform is anticipated to be smaller than the existing state-of-the-art by at least one order of magnitude. Moreover, owing to optimal mode-overlap and negligible coupling losses in bulk crystals, when compared to single mode waveguides, an even higher signal-to-noise ratio improvement is expected.
\par As a caveat, efficient frequency conversion in bulk periodically poled crystals requires higher pump powers due to the absence of waveguide confinement. This can be remedied by using optical cavities to obtain the necessary pump enhancement, while maintaining the low-noise afforded by bulk ppKTP. Such enhancement cavities for bulk crystals have been successfully used in QFC \cite{Albota04,Pan06}, with internal conversion efficiencies of 90\% \cite{Albota04}, still highly competitive among all platforms. 
\par In conclusion, we have demonstrated that the RDC errors that afflict ppLN devices appear substantially reduced in ppKTP. Using an extended SFG technique to sample the phase matching curve at very large detunings, we have bounded the RDC errors for our commercially supplied ppKTP crystals to below 2\% resulting in at least one order of magnitude lower NSDs related to spurious SPDC when compared to current state-of-the-art single-step converters. This promises to solve a challenge that currently severely limits the performance QFC-devices, and consequently, the potential scale of quantum networks.

\section*{Acknowledgments} We thank L. Liebermeister and  B. Globisch from Fraunhofer Institute for Telecommunications HHI for providing equipment.
\section*{Funding Information} Funded by the BMBF, Germany within project Q.Link.X.
\section*{Disclosures}
The authors declare no conflicts of interest.

\end{document}